\documentclass[a4paper,fleqn,usenatbib]{mnras}

\usepackage{newtxtext,newtxmath}

\usepackage[T1]{fontenc}
\usepackage{ae,aecompl}

\usepackage{graphicx}	
\usepackage{amsmath}	
\usepackage{amssymb}	
\usepackage{longtable}

\title[Statistical analysis on X-ray flares in M87 jet]{Statistical analysis on X-ray flares from the nucleus and HST-1 knot in the M87 jet}

\author[Yang et al.]{
Shenbang Yang,$^{1}$
Dahai Yan,$^{2,3}$\thanks{E-mail: yandahai@ynao.ac.cn}
Benzhong Dai,$^{1}$\thanks{E-mail: bzhdai@ynu.edu.cn}
Pengfei Zhang,$^{1}$
Qianqian Zhu,$^{4}$\\
\newauthor
Jiancheng Wang,$^{2,3}$
and Li Zhang$^{1}$
\\
$^{1}$Key Laboratory of Astroparticle Physics of Yunnan Province, Yunnan University, Kunming 650091, China\\
$^{2}$Key Laboratory for the Structure and Evolution of Celestial Objects, Yunnan Observatory,Chinese Academy of Sciences, Kunming
650011, China\\
$^{3}$Center for Astronomical Mega-Science, Chinese Academy of Sciences, 20A Datun Road, Chaoyang District, Beijing 100012, China\\
$^{4}$Department of General Studies, Nanchang Institute of Science \& Technology, Nanchang 330108, China
}

\date{Accepted XXX. Received YYY; in original form ZZZ}

\pubyear{2019}

\begin{document}
\label{firstpage}
\pagerange{\pageref{firstpage}--\pageref{lastpage}}
\maketitle

\begin{abstract}
The statistical properties of X-ray flares from two separate locations (nucleus and HST-1) in the M87 jet are investigated to reveal the physical origin of the flares.
We analyse the archival \textit{Chandra} data for M87, and identify 14 flares in the nucleus and 9 flares in HST-1.
The peak intensity ($I_{\rm{P}}$) and the flaring duration time ($T_{\rm{fl}}$) for each flare are obtained.
It is found that the distributions of both $I_{\rm{P}}$ and $T_{\rm{fl}}$ for the nucleus obey a power-law form with a similar index.
A similar result is also obtained for HST-1, and no significant inconsistency between the nucleus and HST-1 is found for the indices.
Similar to solar X-ray flares, the power-law distributions of the flare event parameters can be well explained by a self-organized criticality (SOC) system,
which are triggered by magnetic reconnection.
Our results suggest that the flares from nucleus and HST-1 are possibly triggered by magnetic reconnection process.
The consistent indices for the distributions of $I_{\rm{P}}$ and $T_{\rm{fl}}$ in the CORE and HST-1 indicate 
that the dimensions of the energy dissipation of the magnetic reconnection is identical in the two regions.
A strong correlation between the flares in the two regions also suggests a similar physical origin for the flares.
\end{abstract}

\begin{keywords}
galaxies: jets -- X-rays: individuals: M87 -- radiation mechanisms: non-thermal
\end{keywords}

\section{Introduction}

Active galactic nucleus (AGN) is believed to host a supermassive black hole (SMBH), which might be the central engine powering a powerful jet.
M87 is a radio galaxy with a central SMBH of about $3 - 6 \times 10^9  $M$_{\sun}$ \citep[e.g.,][]{1997ApJ...489..579M,2009ApJ...700.1690G}
and a collimated jet misaligned by $\sim 30^{\circ}$ with respect to the line of sight \citep[e.g.,][]{1995ApJ...447..582B}.
The nucleus (hereafter, CORE) and several knots in the jet are exposed at radio, optical, and X-ray wavelengths.
Apart from the CORE, the most interesting X-ray emission region is the inner knot HST-1, which is $\sim 1\arcsec$ from the CORE, and was
first discussed by \citet{1999ApJ...520..621B} for the optical region.

Thanks to the unprecedented resolution of \textit{Chandra}, which is $\leqslant 0\farcs5$ \citep{2000SPIE.4012....2W},
the high energy radiation mechanisms in substructures in the M87 jet have been studied in detail.
The X-ray emissions from the CORE and HST-1 are believed to be dominated by non-thermal synchrotron radiation \citep[e.g.,][]{2002ApJ...564..683M,2002ApJ...568..133W,2010ApJ...710.1017Z,2018A&A...612A.106S}. 
Strong flares from the two regions are detected, and the variability timescales span from months to years \citep{2003ApJ...586L..41H,2003ApJ...599L..65P,2006ApJ...640..211H}.

The physical origin of the X-ray flares is not well understood.
Rather than the conventional view that jet emissions are powered by shocks, \citet{2015MNRAS.450..183S} suggested that the magnetic reconnection process powers jet emissions.
Numerical simulations have been performed to test this mechanism, and the results indicated that the variability in AGNs or $\gamma$-ray bursts (GRBs) can be explained with the fast magnetic reconnection process driven by kink-instability turbulence \citep{2016ApJ...824...48S}.

The flares triggered by magnetic reconnection are presumed to form a self-organized criticality (SOC) system, 
such as solar flares \citep[e.g.,][]{1991ApJ...380L..89L,2011SoPh..274...99A}. 
The SOC system predicts that the event parameters, including peak flux and flaring time duration, 
follow power-law distributions, and the corresponding indices are related to the effective geometric dimension of the system \citep[e.g.,][]{2012A&A...539A...2A}.
The SOC model has been proposed to explain the statistical properties of the X-ray flares from GRBs \citep{2013NatPh...9..465W,Yi}, blazars \citep{Yan2018}, 
M87 \citep{2015ApJS..216....8W}, and Sgr A$^\ast$ \citep{2015ApJS..216....8W,2015ApJ...810...19L,2018MNRAS.473..306Y}.

By employing \textit{Chandra} data, \citet{2015ApJS..216....8W} investigated the statistical properties of 18 flares from the nucleus of the M87 jet,
and claimed that those properties can be explained by an SOC model.
Besides the nucleus, HST-1 is another bright X-ray emission region in the M87 jet.
The \textit{Chandra} observations facilitate the study of X-ray flares from different regions in the M87 jet.

We focus on the X-ray flares from the nucleus and HST-1 in the M87 jet,
exploring the statistical properties of the X-ray flares from the two regions.
This enables us to make a comparison between the energy dissipation mechanisms in the two regions.
This paper is organized as following: the \textit{Chandra} data reduction process is described in Sect. \ref{sec:data},
and the results are presented in Sect. \ref{sec:result}. The a discussion and conclusions are provided in Sect. \ref{sec:dis}.

\section{Data reduction and analysis}\label{sec:data}
Until January 2019, M87 has been observed 122 times by \textit{Chandra}. 
The data of 120 archival observations has been publicly released, 
and the observations are obtained by using the Advanced CCD Imaging Spectrometer (ACIS) with a 0.4s/3.2s frame time 
and using the High Resolution Camera (HRC).
To minimize pile-up effects, we reprocessed 105 archival observations with a 0.4 s frame time \citep[e.g.,][]{2006ApJ...640..211H} from the \textit{Chandra} Data Archive (CDA)\footnote{\url{http://asc.harvard.edu/cda/}}.
These observations are all imaged by ACIS. The information for these observations are listed in Table \ref{tab:NO1}.
We perform the data reduction procedures using \textit{Chandra} Interactive Analysis of Observations (CIAO, v4.10) and the \textit{Chandra} Calibration Database (CALDB, v4.7.8).

\begin{figure}
	\centering
	\resizebox{\hsize}{!}
	{\includegraphics{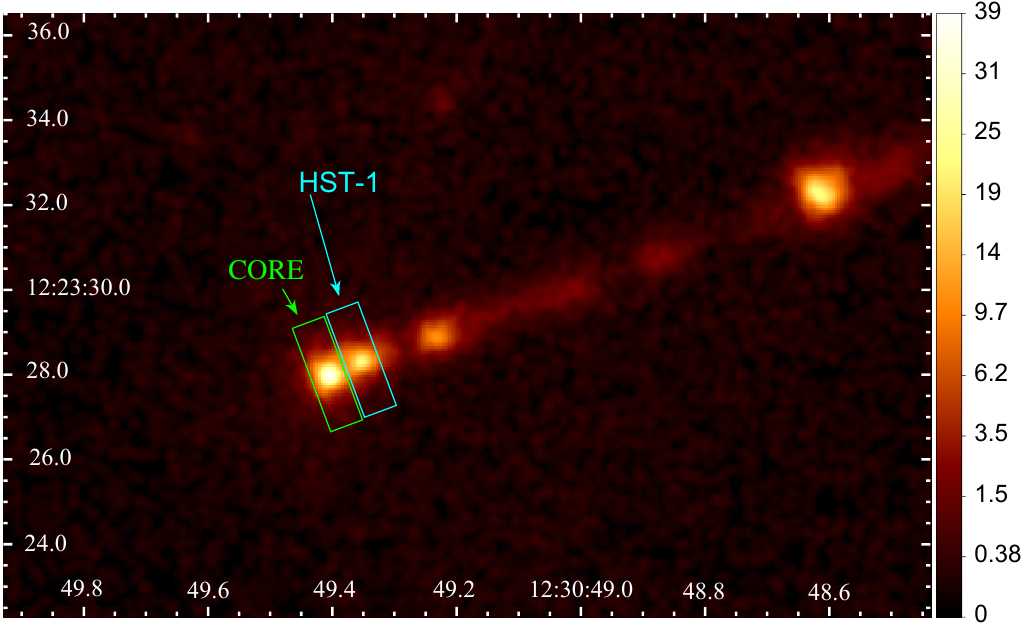}}
	\caption{Image of the observation 1808, binned in $0\farcs123$ per pixel. The x-axis is the right ascension, and the y-axis is the declination. The green and cyan rectangle of length $2\farcs6$ transverse to the jet and 0\farcs8 along the jet are the selected source regions for the CORE and HST-1, respectively, in the data reduction.}\label{1808image}
\end{figure}

However, a series of studies on the M87 jet using \textit{Chandra} data demonstrated that some observations with a 0.4 s frame still suffered from pile-up effects, 
especially for those of HST-1 in an outburst around 2005 \citep{2006ApJ...640..211H,2009ApJ...699..305H,2011ApJ...743..177H}.
The pile-up effect arises when two or more photons are detected by a CCD in a frame time, within a single pixel \citep{2001ApJ...562..575D}.
If an observation of a bright source is affected by the heavy pile-up, the spectrum is distorted so that the original spectral information and integral energy flux can not be obtained.
To recover the intensity of the piled source, \citet{2006ApJ...640..211H} proposed the ``keV s$^{-1}$'' method that extracts the observed intensity of the source in an evt1 file without any grade filtering.
This method is the most efficient way to restore the intrinsic variability of a piled source, 
although some unrecoverable effects, such as ``Eat Thy Neighbor'', ``second-order effects of pileup'' (see \citet{2009ApJ...699..305H} for details), 
can probably induce deviation in the intensity of the source.
We, therefore, use the ``keV s$^{-1}$'' method to estimate the total observed intensities in the CORE and HST-1 following the procedures in \citet{2006ApJ...640..211H}.
For convenience, we use the CIAO tool \texttt{dmstat} to obtain the total energy (in units of eV) of all events on the evt1 files within the rectangular regions of length $2\farcs6$ transverse to the jet and 0\farcs8 along the jet, as shown in Fig. \ref{1808image}.  To separate the CORE from HST-1, \citet{2006ApJ...640..211H} selected a circular region with radius of $0\farcs44$ for HST-1. We extract the events from a rectangle with length 0\farcs8 along the jet (i.e., 0\farcs4 along the jet from the centre of the source) to avoid contamination by the adjacent source.
Henceforth, the intensities can be calculated by {\it total energy/exposure/1000}, and the results are listed in Table \ref{tab:NO1}.
Statistical uncertainties can be calculated based on the counts: $\sqrt{N}/N$, and they are negligible in our analyses since they range from 1 to 5\%, according to \citet{2009ApJ...699..305H}. Only the standard good-time intervals are applied to the evt1 files.

\begin{figure*}
	\centering

		\includegraphics[scale=0.3]{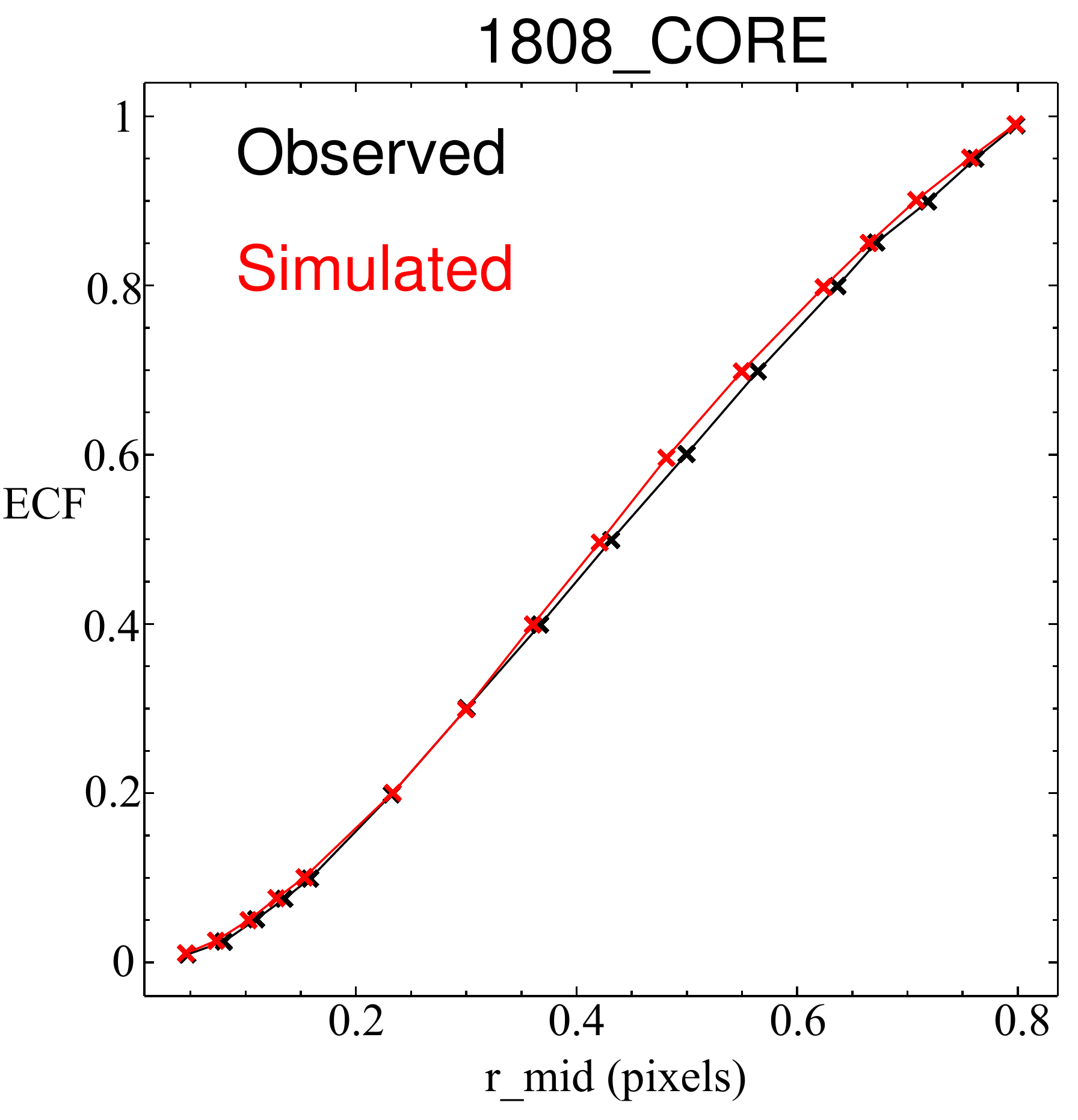}
		\includegraphics[scale=0.3]{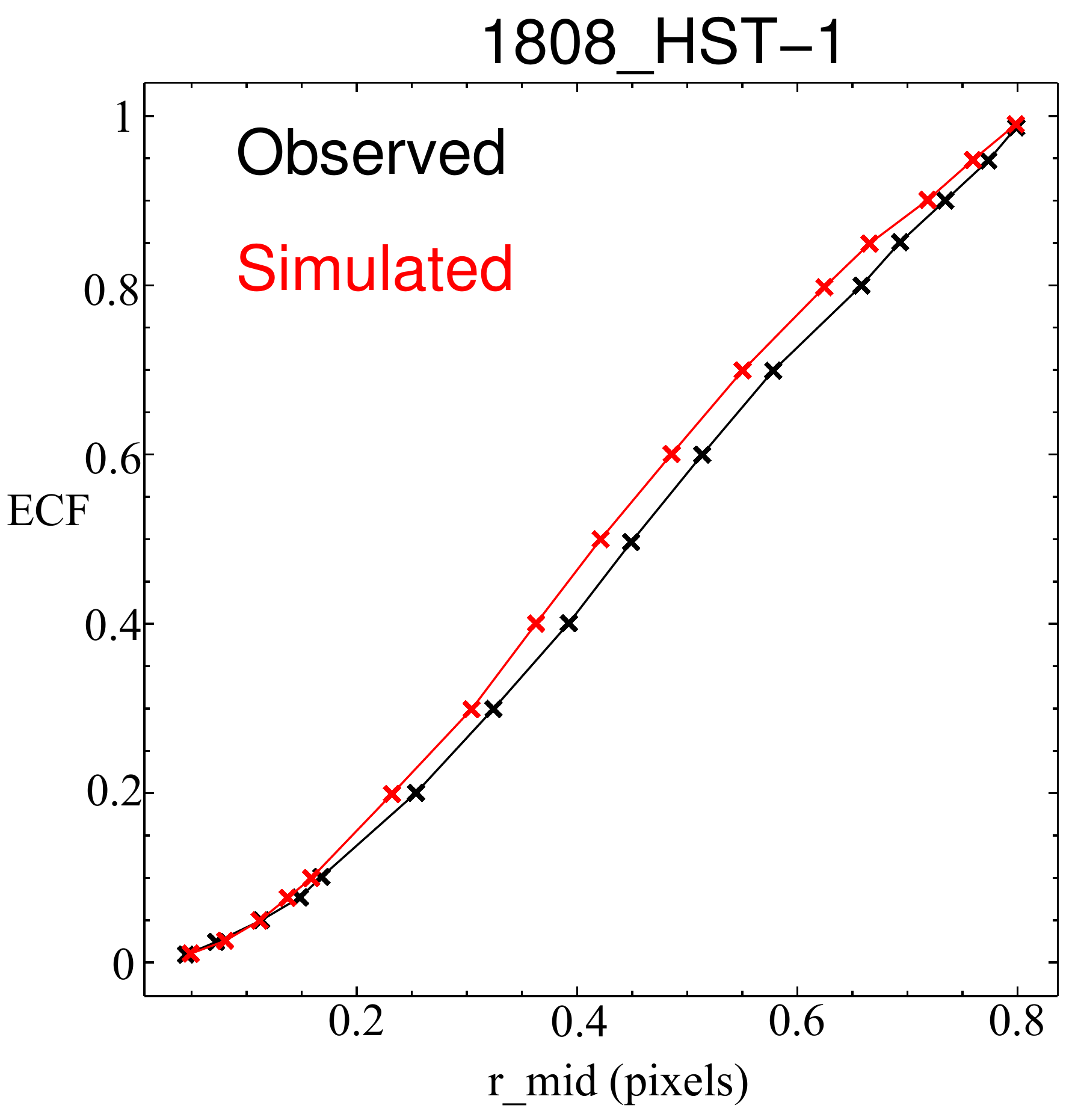}

	\caption{The observed and simulated ECF of the CORE and HST-1 in observation 1808. Variable r\_mid is the radius from the coordinate of the corresponding source.}
	\label{fig:ecf}
\end{figure*}

Furthermore, we examine the possible contamination between the two regions with the point spread function (PSF) simulations, 
using the software MARX\footnote{\url{https://space.mit.edu/ASC/MARX/}}. We first calculate the encircled counts fraction (ECF) within a circle of $0\farcs4$ radius for the CORE and HST-1 in observation 1808.
Then we separately simulate the CORE and HST-1 for the same observation with MARX, and then calculate their ECF within same radius circle.
The results are displayed in Figure \ref{fig:ecf}. If the source in the region of $0\farcs4$ radius is contaminated by the adjacent source, the observed ECF of this source will be distorted.
An apparent discrepancy will then be revealed between the observed and simulated ECF in this case. 
Nevertheless, it has been found that there is no obvious discrepancy in our results, and the contamination is negligible.

\clearpage
\onecolumn
	\begin{longtable}{cccccc}
		\caption{Observational information for M87 and the results of data analysis. (1) Observation ID, (2,3) the start time of the observation, (4) total exposure, (5,6) observed intensity of the CORE and HST-1, respectively, in units of keV s$^{-1}$. It is extracted from the regions shown in Fig. \ref{1808image}. No background is subtracted.\label{tab:NO1}}\\
		\hline\hline
		Obs.ID & \multicolumn{2}{c}{Time}  & Expo. & the CORE & HST-1 \\
		\cline{2-3}
		&UT &MJD & (ks) & $I$ & $I$  \\
		(1)& (2) & (3) & (4) & (5) &(6) \\		
		\hline
		\endfirsthead
		\caption{continued.}\\
		\hline\hline
		Obs.ID & \multicolumn{2}{c}{Time}  & Expo. & the CORE & HST-1 \\
		\cline{2-3}
		&UT &MJD & (ks) & $I$ & $I$  \\
		(1)& (2) & (3) & (4) & (5) &(6) \\		
		\hline		
		\endhead
		\hline
		\endfoot				
		1808	&	2000-07-30T20:42:04	&	51755.8626	&	12.85	&	0.33	&	0.19	\\
		3085	&	2002-01-16T01:25:51	&	52290.0596	&	4.89	&	0.69	&	0.62	\\
		3084	&	2002-02-12T09:03:18	&	52317.3773	&	4.66	&	0.59	&	0.5	\\
		3086	&	2002-03-30T07:36:53	&	52363.3173	&	4.62	&	0.66	&	0.43	\\
		3087	&	2002-06-08T19:58:05	&	52433.832	&	4.97	&	0.47	&	0.74	\\
		3088	&	2002-07-24T14:20:38	&	52479.5977	&	4.71	&	0.58	&	0.99	\\
		3975	&	2002-11-17T02:25:39	&	52595.1012	&	5.29	&	0.7	&	0.7	\\
		3976	&	2002-12-29T14:13:52	&	52637.593	&	4.79	&	0.67	&	0.58	\\
		3977	&	2003-02-04T14:22:12	&	52674.5988	&	5.28	&	0.62	&	0.57	\\
		3978	&	2003-03-09T21:20:35	&	52707.8893	&	4.85	&	0.82	&	0.77	\\
		3979	&	2003-04-14T05:46:54	&	52743.2409	&	4.49	&	0.64	&	0.99	\\
		3980	&	2003-05-18T22:11:43	&	52777.9248	&	4.79	&	0.42	&	0.97	\\
		3981	&	2003-07-03T09:13:36	&	52823.3845	&	4.68	&	0.44	&	0.84	\\
		3982	&	2003-08-08T04:28:37	&	52859.1866	&	4.84	&	0.31	&	1.25	\\
		4917	&	2003-11-11T19:46:06	&	52954.8237	&	5.03	&	0.91	&	2.01	\\
		4918	&	2003-12-29T09:45:40	&	53002.4067	&	4.68	&	0.51	&	1.91	\\
		4919	&	2004-02-12T05:27:42	&	53047.2276	&	4.7	&	0.83	&	3.85	\\
		4921	&	2004-05-13T03:40:04	&	53138.1528	&	5.25	&	0.85	&	4.87	\\
		4922	&	2004-06-23T18:26:35	&	53179.7685	&	4.54	&	0.43	&	5.48	\\
		4923	&	2004-08-05T08:11:51	&	53222.3416	&	4.63	&	0.56	&	5.66	\\
		5737	&	2004-11-26T20:19:16	&	53335.8467	&	4.21	&	1.46	&	6.82	\\
		5738	&	2005-01-24T01:42:58	&	53394.0715	&	4.67	&	1.55	&	7.69	\\
		5739	&	2005-02-14T18:44:29	&	53415.7809	&	5.15	&	1.75	&	8.11	\\
		5740	&	2005-04-22T16:20:31	&	53482.6809	&	4.7	&	1.2	&	11.5	\\
		5744	&	2005-04-28T14:01:45	&	53488.5846	&	4.7	&	1.04	&	11.87	\\
		5745	&	2005-05-04T22:58:04	&	53494.957	&	4.71	&	1.18	&	11.15	\\
		5746	&	2005-05-13T01:51:28	&	53503.0774	&	5.14	&	1.07	&	11.2	\\
		5747	&	2005-05-22T03:20:36	&	53512.1393	&	4.7	&	0.94	&	11.09	\\
		5748	&	2005-05-30T02:31:18	&	53520.1051	&	4.7	&	0.86	&	10.08	\\
		5741	&	2005-06-03T03:31:43	&	53524.147	&	4.7	&	0.86	&	9.39	\\
		5742	&	2005-06-21T01:39:21	&	53542.069	&	4.7	&	0.71	&	9.78	\\
		5743	&	2005-08-06T17:33:28	&	53588.7316	&	4.67	&	0.39	&	7.14	\\
		6299	&	2005-11-29T02:01:54	&	53703.0847	&	4.66	&	0.68	&	3.65	\\
		6300	&	2006-01-05T03:08:49	&	53740.1311	&	4.66	&	0.8	&	3.55	\\
		6301	&	2006-02-19T23:36:31	&	53785.9837	&	4.34	&	1.05	&	3.4	\\
		6302	&	2006-03-30T09:25:20	&	53824.3926	&	4.7	&	0.52	&	3.8	\\
		6303	&	2006-05-21T16:22:42	&	53876.6824	&	4.7	&	0.55	&	3.05	\\
		6304	&	2006-06-28T14:07:05	&	53914.5883	&	4.68	&	1.36	&	2.37	\\
		6305	&	2006-08-02T20:52:12	&	53949.8696	&	4.65	&	0.98	&	2.04	\\
		7348	&	2006-11-13T16:54:29	&	54052.7045	&	4.54	&	0.83	&	4.11	\\
		7349	&	2007-01-04T02:09:51	&	54104.0902	&	4.68	&	0.85	&	3.55	\\
		7350	&	2007-02-13T06:56:01	&	54144.2889	&	4.66	&	0.96	&	3.59	\\
		8510	&	2007-02-15T09:08:50	&	54146.3811	&	4.7	&	0.8	&	3.48	\\
		8511	&	2007-02-18T22:05:59	&	54149.9208	&	4.7	&	0.71	&	3.64	\\
		8512	&	2007-02-21T23:46:51	&	54152.9909	&	4.7	&	0.87	&	3.29	\\
		8513	&	2007-02-24T03:02:04	&	54155.1264	&	4.7	&	0.94	&	3.31	\\
		8514	&	2007-03-12T11:32:42	&	54171.481	&	4.47	&	0.95	&	3.26	\\
		8515	&	2007-03-14T14:21:22	&	54173.5982	&	4.7	&	0.82	&	3.55	\\
		8516	&	2007-03-19T10:21:35	&	54178.4317	&	4.68	&	1.02	&	3.28	\\
		8517	&	2007-03-22T03:48:57	&	54181.159	&	4.67	&	1.02	&	3.38	\\
		7351	&	2007-03-24T19:24:42	&	54183.8088	&	4.68	&	0.97	&	3.58	\\
		7352	&	2007-05-15T11:01:49	&	54235.4596	&	4.59	&	0.56	&	3.0	\\
		7353	&	2007-06-25T14:10:22	&	54276.5905	&	4.54	&	0.5	&	3.35	\\
		7354	&	2007-07-31T01:30:37	&	54312.0629	&	4.71	&	0.38	&	3.52	\\
		8575	&	2007-11-25T17:03:03	&	54429.7105	&	4.68	&	0.66	&	1.33	\\
		8576	&	2008-01-04T23:07:16	&	54469.9634	&	4.69	&	0.69	&	1.19	\\
		8577	&	2008-02-16T11:30:12	&	54512.4793	&	4.66	&	1.55	&	1.15	\\
		8578	&	2008-04-01T19:38:16	&	54557.8182	&	4.71	&	0.86	&	1.37	\\
		8579	&	2008-05-15T04:26:06	&	54601.1848	&	4.71	&	0.65	&	1.32	\\
		8580	&	2008-06-24T06:27:54	&	54641.2694	&	4.7	&	1.12	&	1.15	\\
		8581	&	2008-08-07T18:46:52	&	54685.7826	&	4.66	&	0.46	&	0.94	\\
		10282	&	2008-11-17T22:34:25	&	54787.9406	&	4.7	&	0.4	&	0.55	\\
		10283	&	2009-01-07T09:12:03	&	54838.3834	&	4.71	&	0.45	&	0.52	\\
		10284	&	2009-02-20T22:38:07	&	54882.9431	&	4.7	&	0.43	&	0.45	\\
		10285	&	2009-04-01T07:15:18	&	54922.3023	&	4.66	&	0.44	&	0.4	\\
		10286	&	2009-05-13T22:43:11	&	54964.9467	&	4.68	&	0.53	&	0.38	\\
		10287	&	2009-06-22T11:23:53	&	55004.4749	&	4.7	&	0.57	&	0.33	\\
		10288	&	2009-12-15T20:12:58	&	55180.8423	&	4.68	&	0.67	&	0.29	\\
		11512	&	2010-04-11T21:14:39	&	55297.8852	&	4.7	&	1.26	&	0.32	\\
		11513	&	2010-04-13T14:16:43	&	55299.595	&	4.7	&	0.69	&	0.32	\\
		11514	&	2010-04-15T20:32:42	&	55301.856	&	4.53	&	0.57	&	0.31	\\
		11515	&	2010-04-17T21:47:42	&	55303.9081	&	4.7	&	0.61	&	0.32	\\
		11516	&	2010-04-20T13:20:52	&	55306.5562	&	4.71	&	0.59	&	0.33	\\
		11517	&	2010-05-05T19:25:21	&	55321.8093	&	4.7	&	0.71	&	0.29	\\
		11518	&	2010-05-09T02:39:49	&	55325.111	&	4.4	&	0.53	&	0.29	\\
		11519	&	2010-05-11T11:17:03	&	55327.4702	&	4.71	&	0.49	&	0.33	\\
		11520	&	2010-05-14T09:04:59	&	55330.3785	&	4.6	&	0.47	&	0.3	\\
		13964	&	2011-12-05T00:00:34	&	55900.0004	&	4.54	&	0.51	&	0.23	\\
		13965	&	2012-02-25T08:14:15	&	55982.3432	&	4.6	&	0.48	&	0.21	\\
		14974	&	2012-12-12T06:48:55	&	56273.284	&	4.6	&	0.43	&	0.14	\\
		14973	&	2013-03-12T04:44:30	&	56363.1976	&	4.4	&	0.44	&	0.16	\\
		16042	&	2013-12-26T15:15:16	&	56652.6356	&	4.62	&	0.31	&	0.13	\\
		16043	&	2014-04-02T14:37:46	&	56749.6096	&	4.6	&	0.52	&	0.13	\\
		17056	&	2014-12-17T18:31:58	&	57008.7722	&	4.6	&	0.37	&	0.1	\\
		17057	&	2015-03-19T19:58:41	&	57100.8324	&	4.6	&	0.49	&	0.1	\\
		18233	&	2016-02-23T12:39:14	&	57441.5273	&	37.25	&	0.2	&	0.07	\\
		18781	&	2016-02-24T10:05:22	&	57442.4204	&	39.52	&	0.2	&	0.07	\\
		18782	&	2016-02-26T00:58:57	&	57444.0409	&	34.07	&	0.23	&	0.07	\\
		18809	&	2016-03-12T04:45:53	&	57459.1985	&	4.52	&	0.23	&	0.06	\\
		18810	&	2016-03-13T07:09:25	&	57460.2982	&	4.6	&	0.21	&	0.08	\\
		18811	&	2016-03-14T13:26:22	&	57461.56	&	4.6	&	0.21	&	0.07	\\
		18812	&	2016-03-16T00:08:39	&	57463.006	&	4.4	&	0.22	&	0.07	\\
		18813	&	2016-03-17T03:27:17	&	57464.144	&	4.6	&	0.22	&	0.08	\\
		18783	&	2016-04-20T08:32:11	&	57498.3557	&	36.11	&	0.18	&	0.08	\\
		18232	&	2016-04-27T05:56:21	&	57505.2475	&	18.2	&	0.22	&	0.09	\\
		18836	&	2016-04-28T01:00:45	&	57506.0422	&	38.91	&	0.22	&	0.08	\\
		18837	&	2016-04-30T23:41:33	&	57508.9872	&	13.67	&	0.17	&	0.07	\\
		18838	&	2016-05-28T23:35:37	&	57536.9831	&	56.29	&	0.16	&	0.07	\\
		18856	&	2016-06-12T12:19:40	&	57551.5137	&	25.46	&	0.15	&	0.07	\\
		19457	&	2017-02-15T11:39:35	&	57799.4858	&	4.6	&	0.26	&	0.08	\\
		19458	&	2017-02-16T09:14:58	&	57800.3854	&	4.58	&	0.22	&	0.07	\\
		20034	&	2017-04-11T23:48:07	&	57854.9918	&	13.12	&	0.46	&	0.09	\\
		20035	&	2017-04-14T02:01:37	&	57857.0845	&	13.12	&	0.36	&	0.09	\\
		21075	&	2018-04-22T00:10:57	&	58230.0076	&	9.13	&	0.67	&	0.13	\\
		21076	&	2018-04-24T13:21:29	&	58232.5566	&	9.05	&	0.76	&	0.14	\\	
	\end{longtable}
\clearpage
\twocolumn
\section{Results } \label{sec:result}
\begin{figure*}
	\centering
	\resizebox{\hsize}{!}
	{\includegraphics{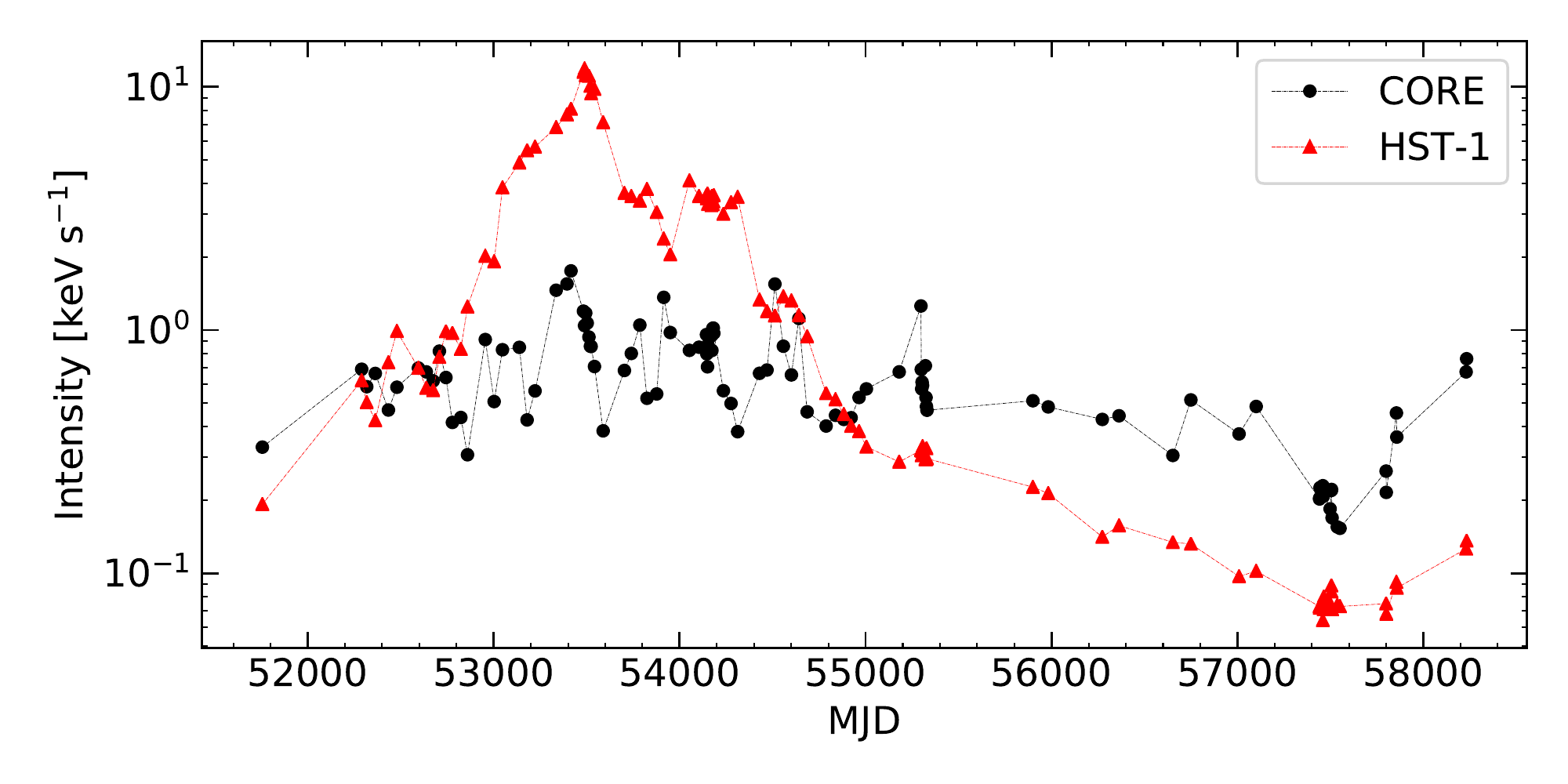}	}
	\caption{The light curves of the CORE (black filled circles) and HST-1 (red filled triangles). The data points are extracted from the regions shown in Fig. \ref{1808image} and listed in Table \ref{tab:NO1}.}
	\label{fig:CORE_lc}
\end{figure*}

The light curves of the CORE and HST-1 are displayed in Figure \ref{fig:CORE_lc}.
The intensities are similar to previous works \citep[e.g.,][]{2006ApJ...640..211H} and \citet{2009ApJ...699..305H}.

\subsection{Distributions of peak intensity and flaring time duration}
We define a criteria that a ``true'' flare should experience a doubling or a halving of intensity, at least.
This criterion could improve the significance of our analysis via filtering fake flares, which are produced by systematic uncertainties.
Practically, we first mark the maximums of peaks and the minimums between neighbouring peaks in the light curves.
Thereafter, we divide each peak intensity by the left and right adjacent minimums.
When a value of $ \gtrsim2 $ ($ \geqslant1.95 $ in practice) is obtained, a ``true'' flare is identified.
Ultimately, 14 flares for the CORE and 9 flares for HST-1 are identified through this procedure, and are listed in Table \ref{tab:NO2}.
We mark the flares identified only by the rising part with the character ``r'', and the flares identified only by the declining part with the character ``d''. 
The flares identified by both parts are labeled with the character ``m''.
All peak intensities ($I_{\rm p}$) of the flares can be directly read out. 
The flaring time durations ($T_{\rm fl}$) are calculated using the definition of $T_{\rm fl} = \rm EndTime-StartTime$.


Two models, a power-law model and a log-normal model, are used to fit the data.
The reliable method for the estimation of the parameters of a statistic distribution model is Maximum likelihood (ML).

For a power-law distribution $p(x)=A(\alpha, x_{\rm{max}},x_{\rm{min}})\cdot x^{-\alpha}$, its logarithmic likelihood function can be expressed as
\begin{equation}\label{fun:like}
\ln L = -\alpha \sum\limits_{i=1}^{N}\log(x_i)+N\cdot\log(A),
\end{equation} 
where the normalization coefficient $A$ can be written as $A=1-\alpha{x_{{\rm max}}^{1-\alpha}-x_{{\rm min}}^{1-\alpha}}$.
Specifically, when $\alpha=1$, $A=1/\log(x_{\rm{max}}/x_{\rm{min}})$.

For a log-normal distribution, $\mathcal{N}(\log x;\mu,\sigma)$, the logarithmic likelihood function is
\begin{equation}\label{fun:lognormlike}
\ln L = -\frac{N}{2}\log(2\pi\sigma^2)-\sum\limits_{i=1}^{N}\log(x_i)-\dfrac{1}{2\sigma^2}\sum\limits_{i=1}^{N}(\log(x_i)-\mu)^2.
\end{equation} 

We use the  Markov Chain Monte Carlo (MCMC) technique \citep[e.g., ][]{2013ApJ...765..122Y,2015MNRAS.454.1310Y} to maximize the likelihood functions, 
and obtain the best-fit parameters of the power-law and log-normal models.
The results are given in Table \ref{tab:allfit}. 

\begin{figure*}
	\centering
	\resizebox{\hsize}{!}
	{
		\includegraphics{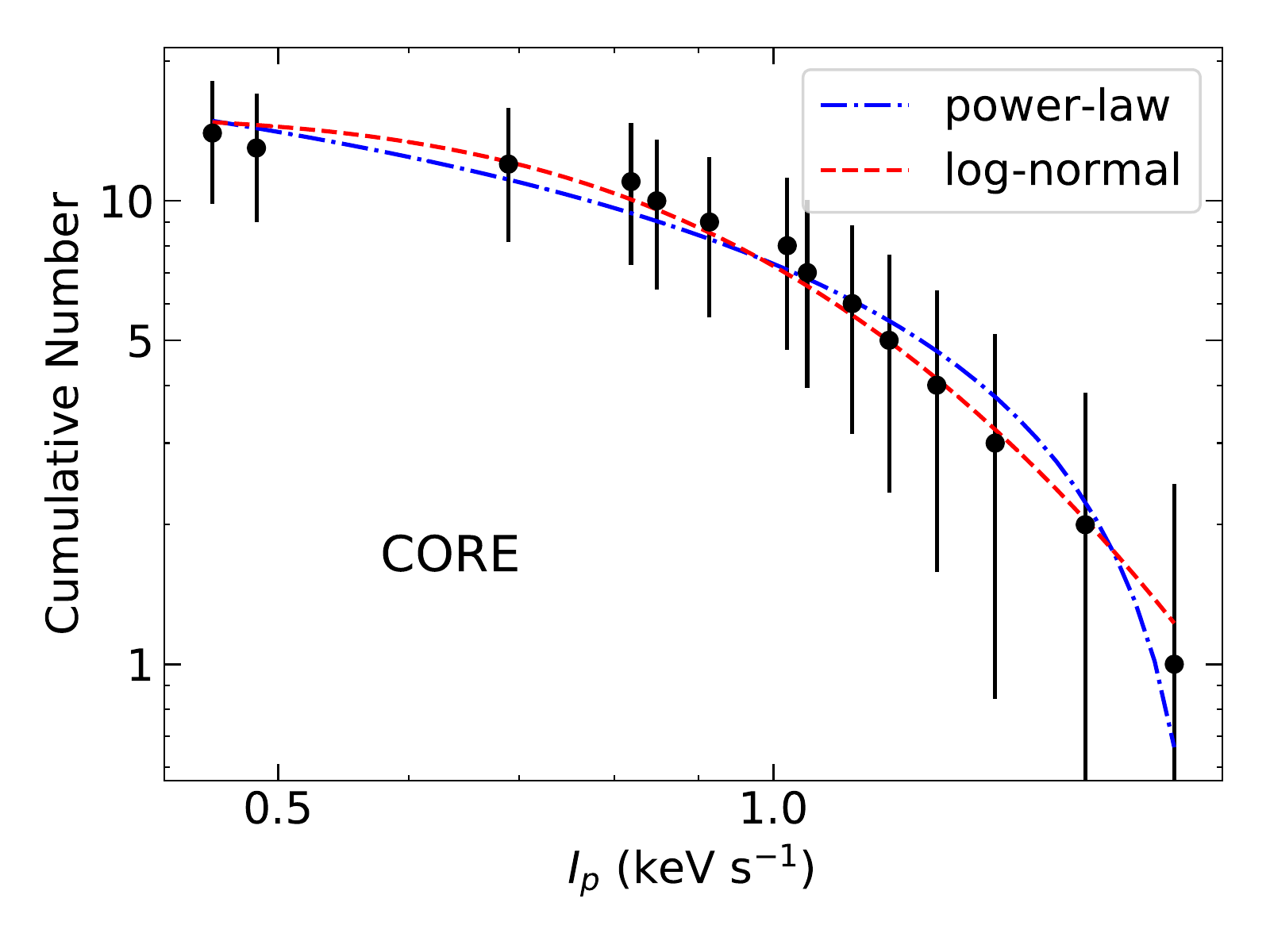}
		\includegraphics{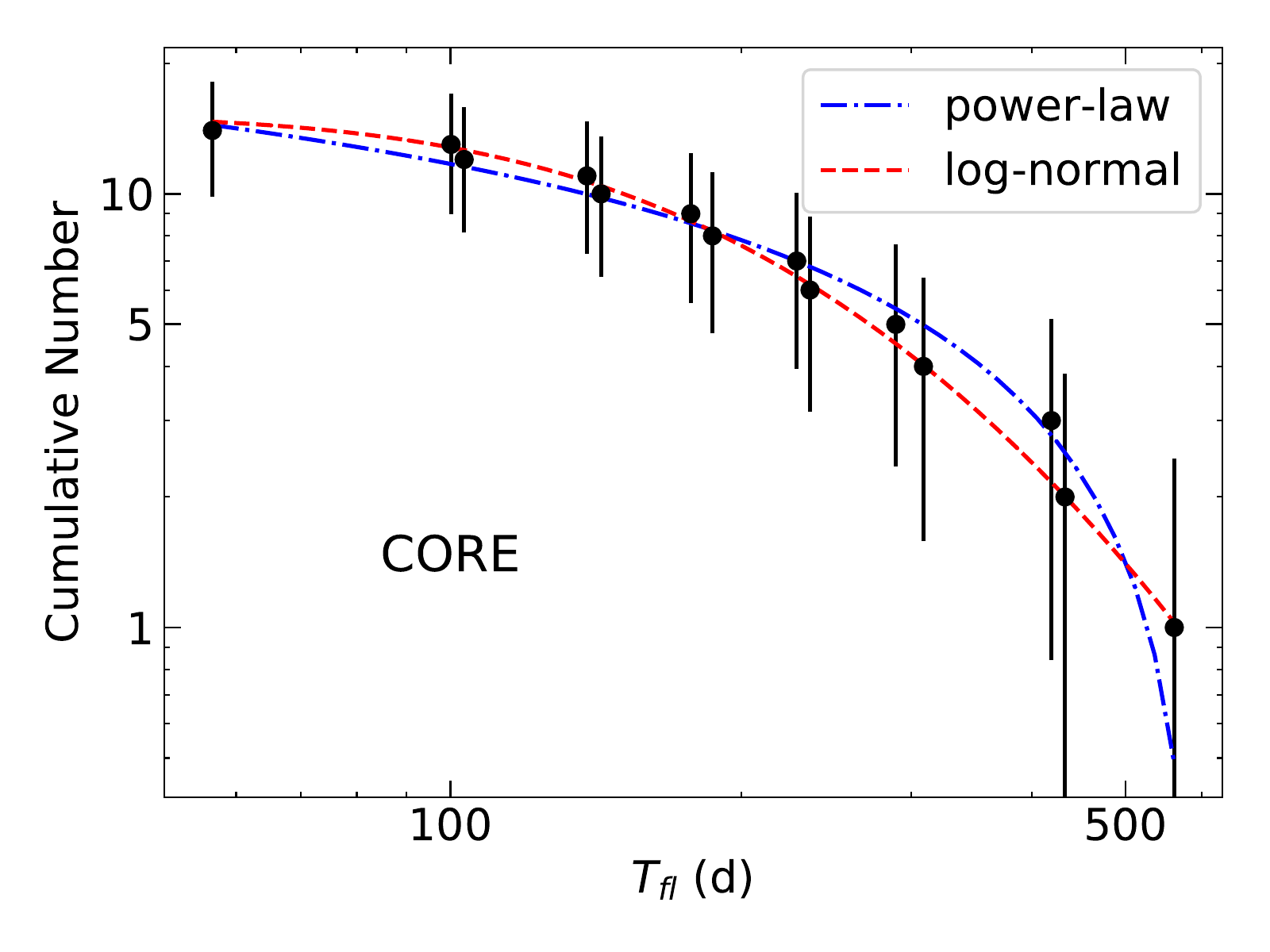}
	}
	\resizebox{\hsize}{!}
	{
		\includegraphics{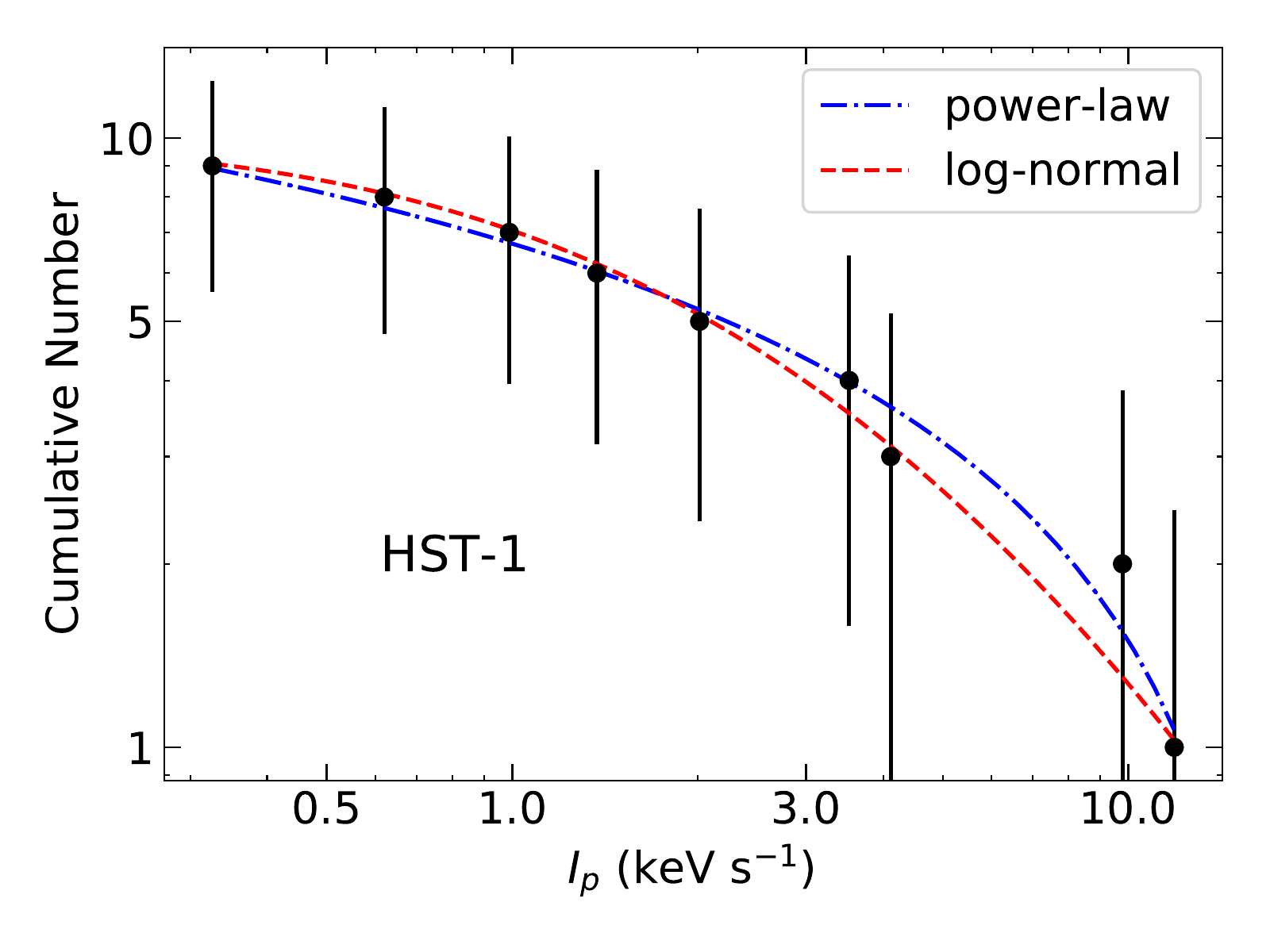}
		\includegraphics{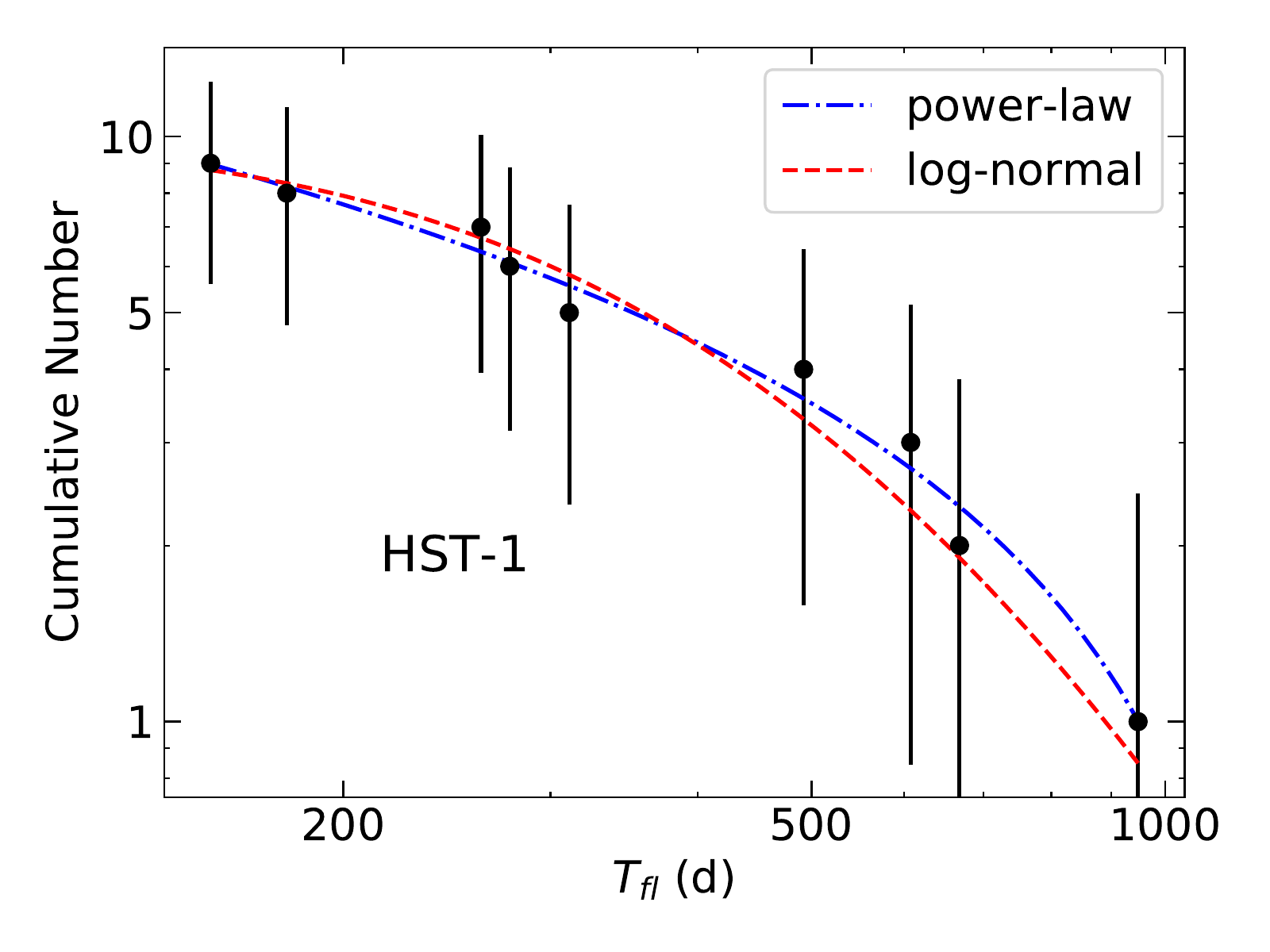}
	}
	\caption{Comparison of the best-fit results and the data $I_{\rm p}$/$T_{\rm{fl}}$. 
	              Errors of the data points are at $1\sigma$ level. The dash-dotted and the dashed lines are the best-fit results of the power-law and log-normal models, respectively. 
	              See the text for details.}
	\label{fig:cum_fit}
\end{figure*}

In Figure \ref{fig:cum_fit}, we show the comparison of the data points and our best-fit results.
Given the small number of the flares, we adopt the cumulative distribution.
The results of the power-law model (dot-dashed lines) are calculated with $N_{\rm PL}(>x)$ and the corresponding best-fit parameters, 
where 
\begin{equation}\label{fun:cdf}
N_{\rm PL}(>x) = \dfrac{a}{\alpha-1}(x^{1-\alpha}-x_{\rm{max}}^{1-\alpha})+b ,
\end{equation} 
where $a$ and $b$ are normalization parameters, $x_{\rm{max}}$ is the maximum cutoff parameter. 
The results of the log-normal model (dashed lines) are calculated with $N_{\rm LN}(>x)$ and the corresponding best-fit parameters,
where
\begin{equation}\label{fun:cum_log}
N_{\rm LN}(>x)=H \cdotp \left(\frac{1}{2}-\frac{1}{2} {\rm erf} \left( \dfrac{\log  x - \mu}{\sqrt{2}\sigma}\right) \right),
\end{equation}
where $H$ is a normalization parameter, erf is the error function. 
One can see that both the power-law and log-normal models match the data well.


The statistical criterion for model selection is  Akaike information criterion (AIC) or Bayesian information criterion (BIC). 
Generally, AIC and BIC are respectively written as ${\rm AIC}=2k+C$ and ${\rm BIC}=k\ln(n)+C$, 
where $k$ is the number of model parameters, $n$ is the number of data, and $C\equiv-2\ln L$. 
In our case, AIC is used as the criterion for the model selection, 
because $n$ is identical in the fits of the models. 
As shown in Table \ref{tab:allfit}, the AIC values for the power-law model are smaller than those for the log-normal model, which means that the power-law model is better.

\begin{figure*}
	\centering
	\resizebox{\hsize}{!}
	{
		\includegraphics{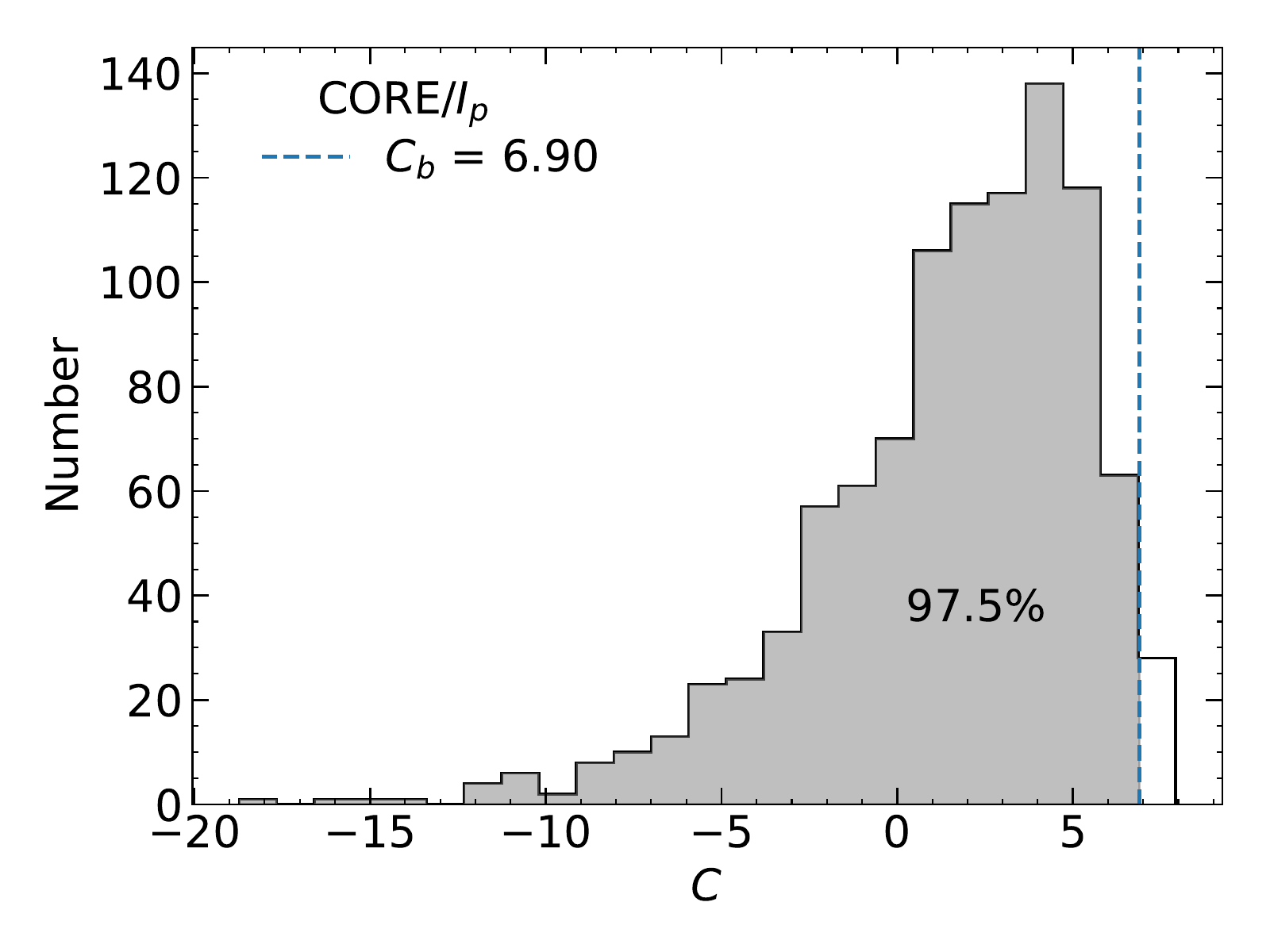}
		\includegraphics{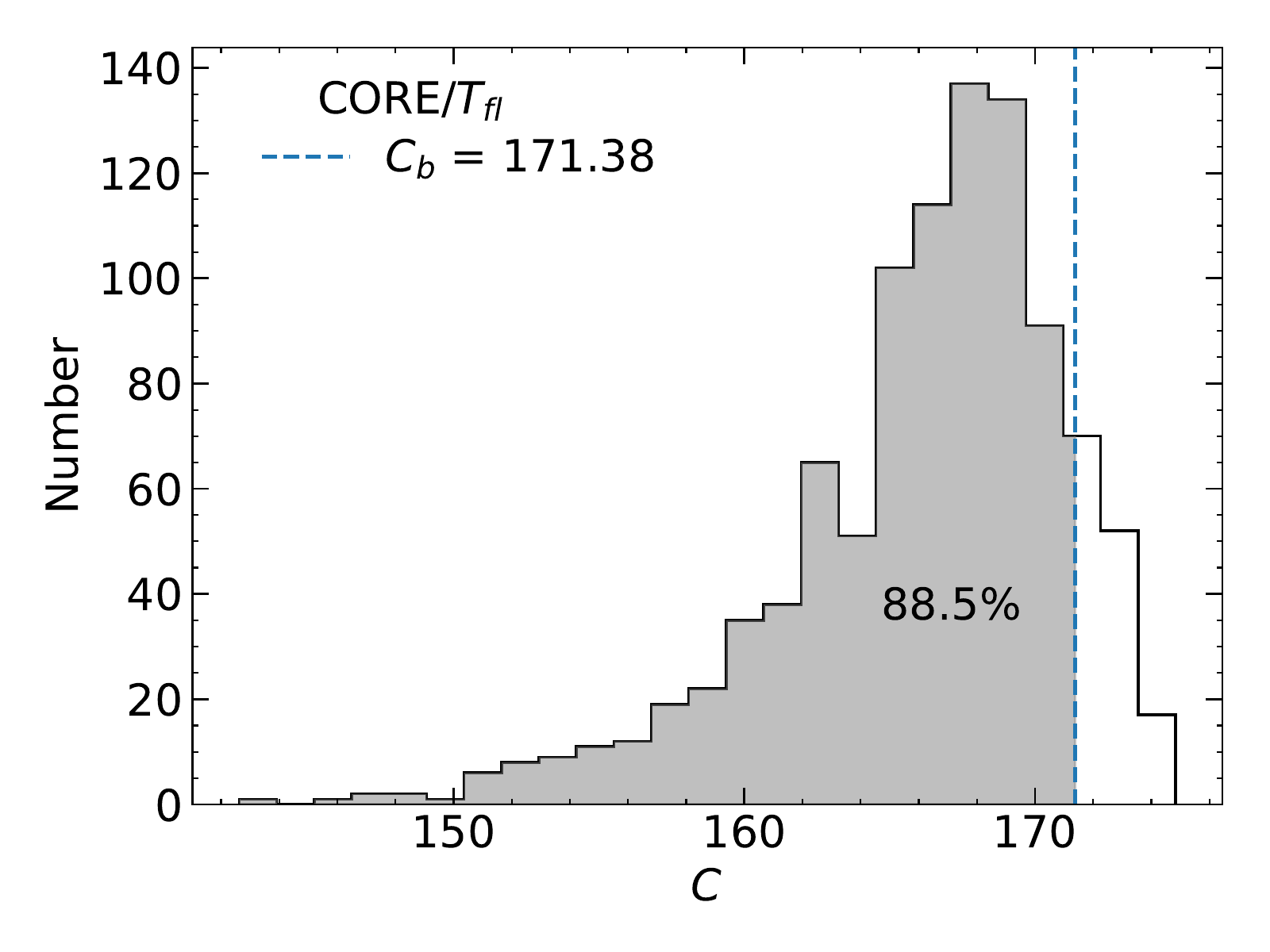}
	}
	\resizebox{\hsize}{!}
	{
		\includegraphics{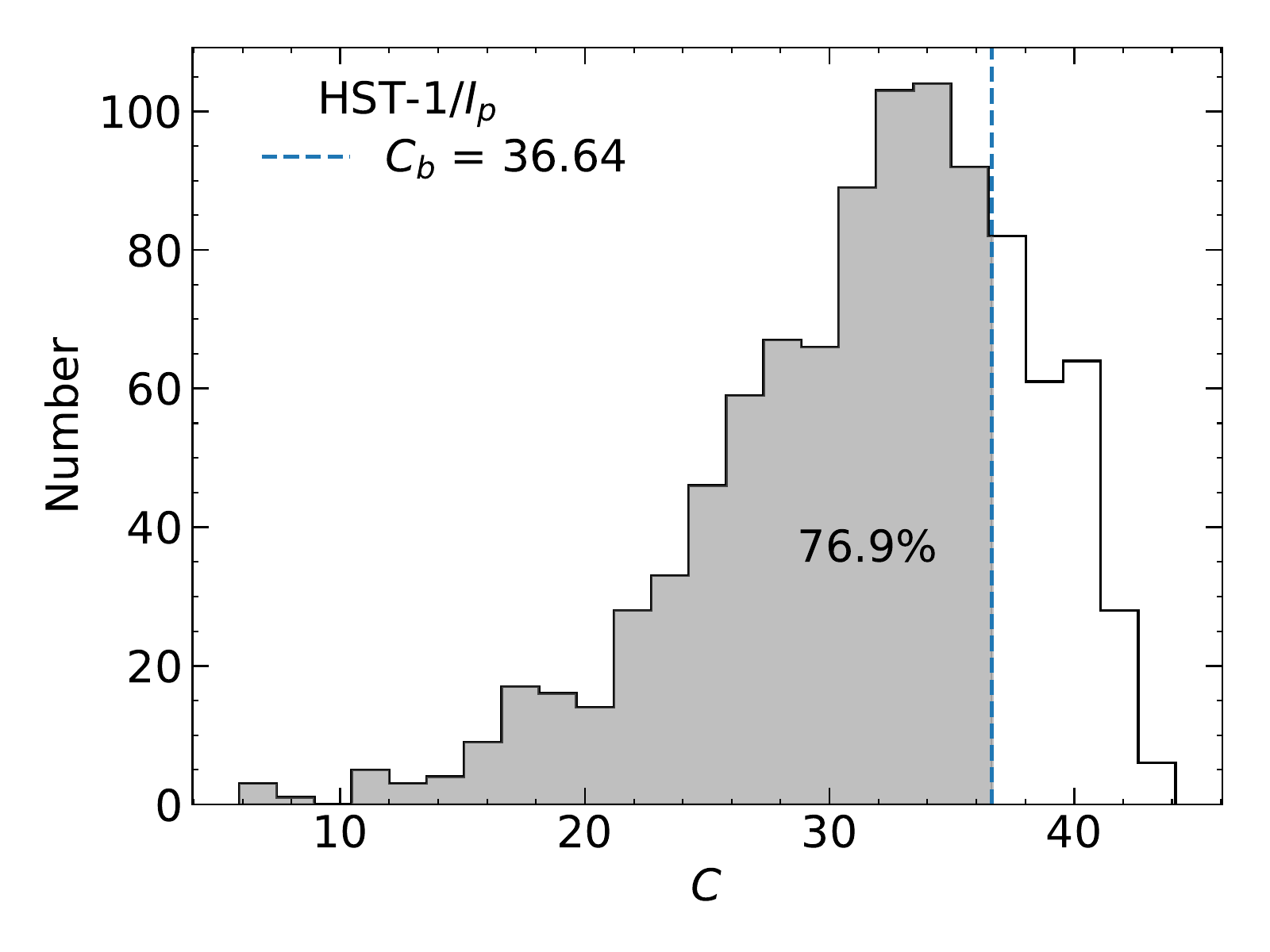}
		\includegraphics{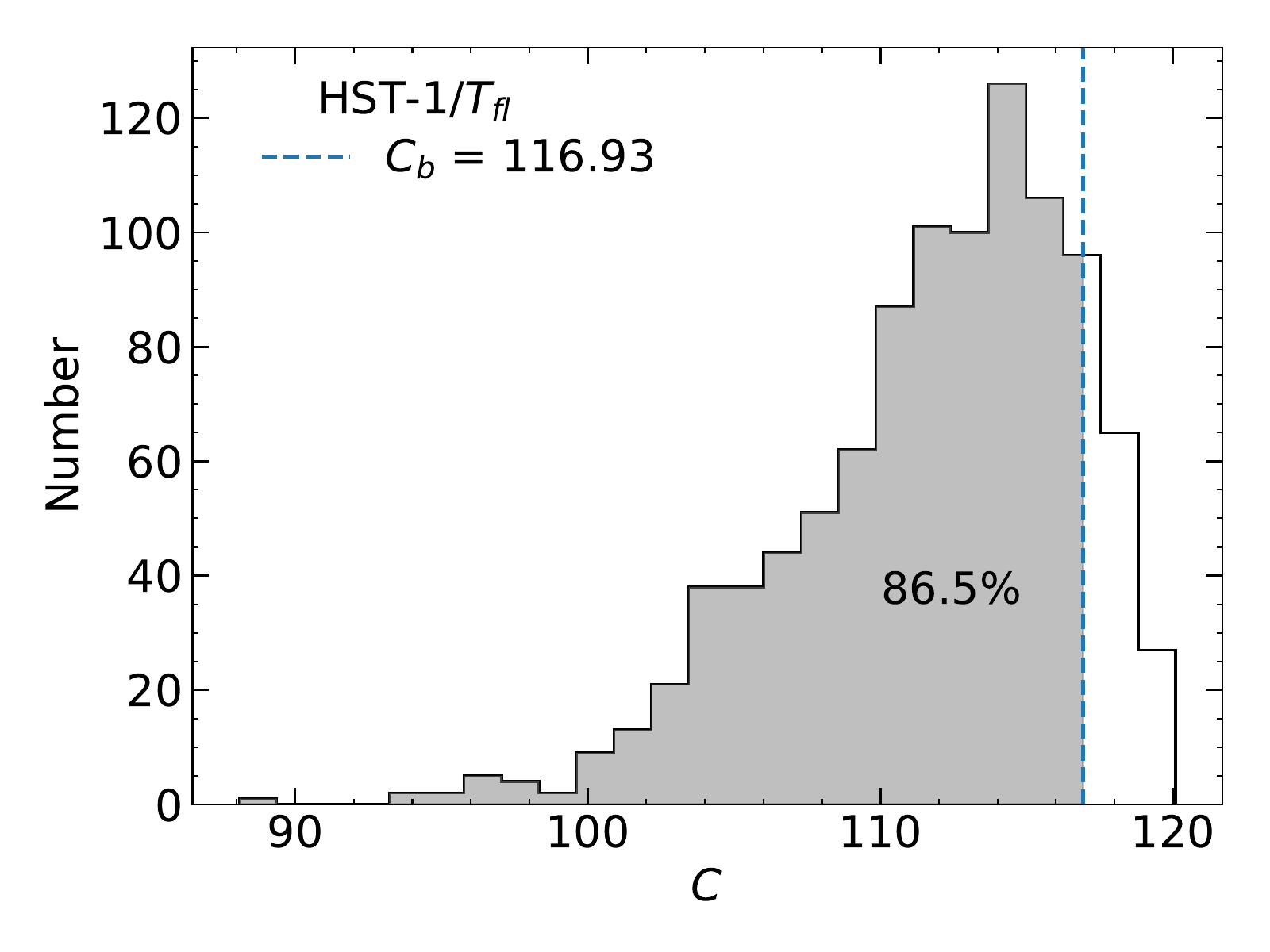}
	}
	\caption{The goodness-of-fit for the best-fits of the power-law model. 
	Distributions of the $C$ (defined as $C\equiv-2\ln L$) value are calculated from the best-fits of the 1000 simulated data. 
	$C_{\rm{b}}$ is the best-fit value for the real data.}
	\label{fig:gof}
\end{figure*}

To access the goodness-of-fit for the power-law model, 
we perform the parametric bootstrap method for the power-law model \citep[e.g.,][]{2018MNRAS.473..306Y}. 
We simulate 1000 data sets using the best-fit parameters of the power-law model, 
and obtain the best-fit values of $C$ from those simulated data. 
The results are shown in Figure \ref{fig:gof}. 
One can see that the number fraction with $C$ smaller than that
of the actual data is $<$ 95\% for the distributions of CORE/$T_{\rm{fl}}$, HST-1/$T_{\rm{fl}}$ and HST-1/$I_{\rm{p}}$. 
This suggests that the three distributions are well described by the power-law model.
While the fraction is 97.5\% for CORE/$I_{\rm{p}}$ distribution,  which means
a worse fitting.
	
\begin{table*}
	\centering
	\caption{Identified flares of the CORE and HST-1. The start, peak, and end times are MJD. The peak intensity ($I_{\rm{p}}$) is in units of keV s$^{-1}$. The Col. Type demonstrates the specified parts that are used to identify the flares. See the text for the details of the identification procedures and the explanations for the symbols.}
	\label{tab:NO2}
	\begin{tabular}{ccccc|ccccc}
		\hline\hline
		\multicolumn{5}{c|}{CORE} & \multicolumn{5}{c}{HST-1} \\
		\hline
		StartTime & PeakTime & EndTime &$I_{\rm{p}}$	& \multicolumn{1}{c|}{Type} &StartTime& PeakTime & EndTime	& $I_{\rm{p}}$ & Type\\		
		\hline
		51755.8626	&	52290.0596	&	52317.3773	&	0.69	&	r	&	51755.8626	&	52290.0596	&	52363.3173	&	0.62	&	r	\\
		52859.1866	&	52954.8237	&	53002.4067	&	0.914	&	r	&	52363.3173	&	52479.5977	&	52674.5988	&	0.989	&	r	\\
		53179.7685	&	53415.7809	&	53488.5846	&	1.751	&	r	&	52823.3845	&	52954.8237	&	53002.4067	&	2.014	&	r	\\
		53824.3926	&	53914.5883	&	54052.7045	&	1.363	&	r	&	53002.4067	&	53488.5846	&	53494.957	&	11.872	&	r	\\
		57800.3854	&	57854.9918	&	57857.0845	&	0.456	&	r	&	53949.8696	&	54052.7045	&	54104.0902	&	4.114	&	r	\\
		52674.5988	&	52707.8893	&	52777.9248	&	0.819	&	d	&	53524.147	&	53542.069	&	53785.9837	&	9.783	&	d	\\
		53002.4067	&	53138.1528	&	53179.7685	&	0.849	&	d	&	54235.4596	&	54312.0629	&	54512.4793	&	3.522	&	d	\\
		53488.5846	&	53494.957	&	53588.7316	&	1.175	&	d	&	54512.4793	&	54557.8182	&	55180.8423	&	1.372	&	d	\\
		54173.5982	&	54181.159	&	54312.0629	&	1.019	&	d	&	55325.111	&	55327.4702	&	56273.284	&	0.326	&	d	\\
		54601.1848	&	54641.2694	&	54787.9406	&	1.116	&	d	&		&		&		&		&		\\
		57008.7722	&	57100.8324	&	57441.5273	&	0.485	&	d	&		&		&		&		&		\\
		53588.7316	&	53785.9837	&	53824.3926	&	1.048	&	m	&		&		&		&		&		\\
		54312.0629	&	54512.4793	&	54601.1848	&	1.546	&	m	&		&		&		&		&		\\
		54882.9431	&	55297.8852	&	55301.856	&	1.256	&	m	&		&		&		&		&		\\
		\hline
	\end{tabular}
\end{table*}
\begin{table*}
	\centering
	
	\caption{The best-fit results of power-law and log-normal. (1) two emission regions, (2) the distribution to be fitted, (3) power-law index , (4) minimum -2$\ln L$ of power-law fit, (5-6) two parameters of log-normal model, (7) minimum -2$\ln L$ of log-normal fit. }\label{tab:allfit}
\begin{tabular}{ccccccc}
	\hline\hline
	& & \multicolumn{2}{c}{power-law} &\multicolumn{3}{c}{log-normal}  \\
	\cline{3-4} \cline{5-7}	
	 &  & $\alpha$ & -2$\ln L$  & $\mu$ & $\sigma$ &  -2$\ln L$  \\	
	(1) & (2) & (3)&(4)&(5)&(6)&(7)\\
	 \hline
	CORE & $T_{\rm{fl}}$ & $0.73_{-0.37}^{+0.39}$ & 171.38& $5.30\pm0.18$ & $0.68_{-0.12}^{+0.17}$ & 175.17 \\
	& $I_{\rm{p}}$ & $0.69_{-0.45}^{+0.59}$ & 6.90 & $-0.03\pm0.11$ & $0.42_{-0.07}^{+0.10}$ & 12.06  \\
	HST-1 & $T_{\rm{fl}}$ & $1.19_{-0.60}^{+0.64}$ & 116.93& $5.90\pm0.24$ & $0.70_{-0.16}^{+0.25}$ & 122.67  \\
	& $I_{\rm{p}}$ & $0.92\pm0.32$ & 36.64 & $0.76\pm0.46$ & $1.36_{-0.30}^{+0.49}$ & 42.13 \\	
	\hline
\end{tabular}
\end{table*}

\subsection{Correlation of light curves}
\begin{figure}
	\centering
	\resizebox{\hsize}{!}
	{\includegraphics{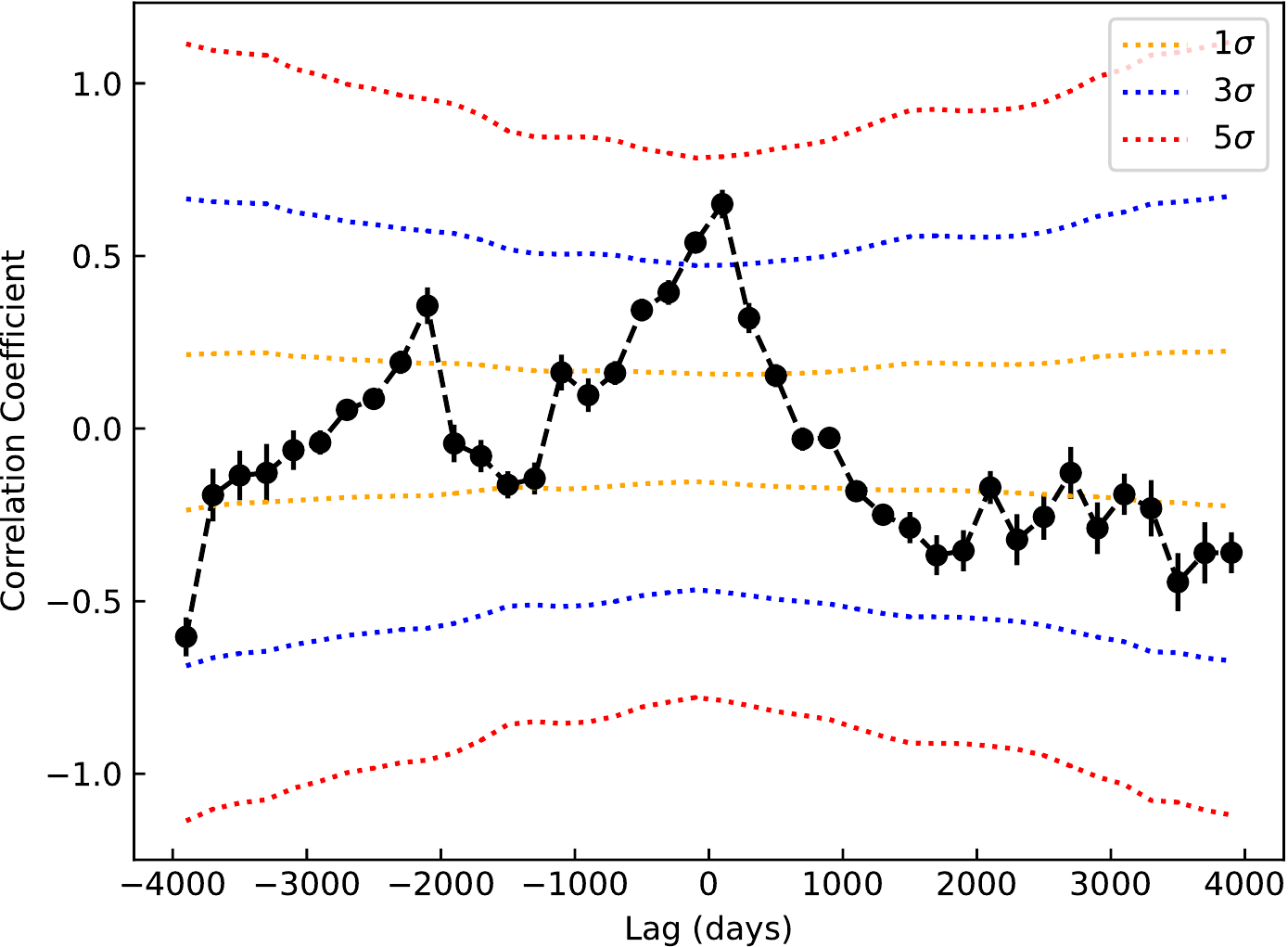}
	}
	\caption{Result of DCF between the intensity light curves of the CORE and HST-1. The bin size of the DCF is 200 days.
		The orange, blue, and red dotted lines correspond to 1, 3, and 5$\sigma$ significance levels, respectively.}
	\label{fig:dcf}
\end{figure}
A frequently used method for the correlation analysis is the discrete correlation function \citep[DCF,][]{1988ApJ...333..646E}.
For two evenly sampled light curves, the significance level of the estimated correlation coefficients by DCF
can be evaluated by the simulation of light curves using the corresponding power spectrum density (PSD).
The light curve simulation technique was proposed by \citet{1995A&A...300..707T} and improved by \citet{2013MNRAS.433..907E}.
However, it is difficult to obtain the correct PSD for an unevenly sampled light curve.
\citet{2014ApJ...788...33K} have proposed a flexible approach to characterize the PSD for irregularly sampled variability.
The light curve can be fitted by using the continuous-time autoregressive moving average (CARMA($p,q$)) model.
Thereafter, the corresponding PSD can be expressed as the sum of Lorentzian functions with the best-fitting parameters.
The significance levels can be estimated for the case of unevenly sampled data with the relatively accurate PSD.
The DCF results are shown in Fig. \ref{fig:dcf}. A strong correlation can be observed among the intensity light curves of the CORE and HST-1, with a significance of $>3\sigma$.

\section{Discussion and conclusions } \label{sec:dis}

Using the \textit{Chandra} observations, we study the statistical properties of the X-ray flares from CORE and HST-1 in M87 jet. 
14 flares and 9 flares are extracted from the CORE and HST-1 light curves, respectively. 
We use the power-law and log-normal models to fit the distributions of $I_{\rm{p}}$ and $T_{\rm{fl}}$ in the two regions. 
The results show that both models can fit the data well. 
While, the statistical criteria suggests that the power-law model is the better one.
We furthermore examine the goodness-of-fit of the power-law model via bootstrapping sampling, and
 the results show that the data can be well characterized by the power-law model.
Note that the fit to the CORE/$I_{\rm{p}}$ distribution by the power-law model is not very well.
More data are needed to examine this distribution.

It has been proposed that the statistical properties of astrophysical X-ray flares from black hole systems can be explained by a fractal-diffusive SOC model
\citep[e.g.,][]{2013NatPh...9..465W,2015ApJS..216....8W,2015ApJ...810...19L,Yan2018}.
The SOC model expects that the distributions of event parameters, such as $I_{\rm{p}}$ and $T_{\rm{fl}}$ follow a power-law distribution.
The corresponding indices can be expressed as $\alpha_{\rm{F}}=1+(S-1)/D_S,$  where $D_S$ is the fractal Hausdorff dimension, spanning from 1 to $S$, and $\alpha_{\rm{T}}=1+\beta(S-1)/S,$ where $\beta$ is a diffusion parameter) \citep[e.g.,][]{2012A&A...539A...2A}. Here $S$ is the spatial dimension of the energy dissipation domain.

\citet{2015ApJS..216....8W} investigated the statistical properties of 18 X-ray flares from the CORE in the M87 jet, and argued that SOC system with $S=3$ could explain the statistical properties.
We find that both the distributions of $I_{\rm{p}}$ and $T_{\rm{fl}}$ in CORE obey a power-law form, 
with the index $\alpha_{\rm{F}}=0.69_{-0.45}^{+0.59}$ and $\alpha_{\rm{T}}=0.73_{-0.37}^{+0.39}$.
our results are consistent with \citet{2015ApJS..216....8W} within the errors.

Furthermore, it is found that both the distributions of $I_{\rm{p}}$ and $T_{\rm{fl}}$ in HST-1 also obey a power-law form,
with $\alpha_{\rm{F}}=0.92\pm0.32$ and $\alpha_{\rm{T}}=1.19_{-0.60}^{+0.64}$.
Each index is consistent with that in the CORE within the errors,.

It is noted that the uncertainties of the power-law indices are large, which is caused by the small number of the flares.
More flares are needed to confirm the present results.

The statistical results of the CORE and HST-1 X-ray flares agree with the SOC model.
This indicates that the X-ray flares in the CORE and HST-1 are possibly triggered by the magnetic reconnection, 
and the energy dissipation dimension was identical in the two regions.
The strong correlation between the light curves in the CORE and HST-1 also hints that the flares in the two regions are triggered by the same mechanism.
Our results support the argument that the magnetic reconnection is a natural process of energy dissipation in relativistic jets \citep[e.g.,][]{2015MNRAS.450..183S,2016ApJ...824...48S}.

Indeed, several models, including the current-driven (CD) kink instability and Kelvin-Helmholtz (KH) instability, have been proposed for the development of complex structures in astrophysical jets \citep{1992SvAL...18..356L,1996ApJ...467..597B,1998ApJ...493..291B}.
For blazars, the CD kink instability is believed to be the one occurring in the relativistic jets \citep{1992SvAL...18..356L,1998ApJ...493..291B,1999MNRAS.308.1006L}.
The relativistic jet of M87 may induce similar CD kink instability, which is considered to be the driving mechanism behind  the magnetic reconnection.
The magnetic fluids suffer a CD kink instability within a short distance from the central engine. Hence, the jet is distorted by the instability, and structures such as knots are formed.
Consequently, the distorted or disrupted magnetic field structures may trigger the magnetic reconnection process \citep[e.g.,][]{2016ApJ...824...48S}.
The energy of the jet will be dissipated by converting the magnetic energy to the kinetic energy of the emitting particles.

\section*{Acknowledgements}
We thank the anonymous referee for the constructive comments and suggestions.  
We acknowledge financial support from National Key R\&D Program of China under grant No. 2018YFA0404204 and the National Natural
Science Foundation of China (NSFC-11803081, NSFC-U1738124, NSFC-11573060, NSFC-11661161010, NSFC-U1531131).
The work of D. H. Yan is also supported by the CAS ``Light of West China'' Program and Youth Innovation Promotion Association.
S. Yang acknowledges the support from Yunnan Provincial Education Office Scientific Research Foundation Project (2019Y0023).

\bibliographystyle{mnras}
\bibliography{m87} 

\bsp	
\label{lastpage}
\end{document}